\documentstyle[aps,axodraw]{revtex}
\tighten
\newcommand{\be}{\begin{equation}}
\newcommand{\ee}{\end{equation}}
\newcommand{\br}{\begin{eqnarray}}
\newcommand{\er}{\end{eqnarray}}
\begin{document}

\title{Calculation of density fluctuation in inflationary epoch}
\author{Amit Kundu,~ S. Mallik }
\address{Saha Institute of Nuclear Physics,\\ 1/AF Bidhannagar, 
Calcutta- 700 064, India}
\author{{\it and} \\ D. Rai Chaudhuri}
\address{Presidency College, Physics Department,\\ 87, College Street, 
Calcutta - 700 012, India} 
\maketitle
\vspace{1cm}

\begin{abstract}
Starting from an initial state of thermal equilibrium, we derive an
expression for the quantum fluctuation in the energy density during the
inflationary epoch in terms of the mode functions for the inflaton field.
The effect of this particular initial state is not washed out in the final
formula, contrary to what is usually believed. Numerically, however,
the effect is completely negligible, validating the use of the two point 
function in the vacuum state. We also point out the requirement of
conventional quantum field theory during inflation, that the quantum 
fluctuation in a wavelength must be evaluated, at the latest, when the 
wavelength crosses the Hubble length, in contrast to the usual 
practice in the literature.

\bigskip
\leftline{PACS number(s): 98.80.Cq}
\end{abstract}
\section{Introduction}
\setcounter{equation}{0}

The most attractive aspect of inflationary models of the 
early universe \cite{guth}
is their potential to predict the present day density 
inhomogeneity from first 
principles \cite{hawking}. In these models it is possible to calculate quantum 
fluctuation in the energy density on the homogeneous background in a region 
within the causal horizon (given by the Hubble length) during the 
inflationary epoch. This fluctuation
provides the initial spectrum of density perturbation. As the region inflates
into the observed universe or bigger, its propagation 
through different eras can be followed till the present time by the 
equations of linear perturbation theory of classical gravity \cite{bardeen}.

In this work we consider some points in the calculation 
of quantum fluctuation during inflation. The basic ingredient is the 
expectation value of the product of two scalar field operators at a time when 
considerable inflation has already taken 
place. It is generally believed that as the inflation proceeds, the 
effects of all scales associated with a particular initial state tend 
to be wiped out, retaining only the extremely high energies associated with 
quantum fluctuations in the vacuum. So the expectation value is evaluated 
for the homogeneous background, which corresponds to the vacuum state 
of quantum field theory.

Here we investigate how far the density fluctuation is actually 
independent of the initial condition prevailing at the beginning of inflation.
For this purpose, we start with a definite initial state, namely that of 
thermal equilibrium. The first attempt in this direction was by Guth and Pi
\cite{gupi}. We discuss it here in a more general framework \cite{semenoff},
\cite{leutwyler}. The thermal 
propagator can be followed till the time when the quantum fluctuations are 
evaluated.

The existence of an initial thermal equilibrium distribution 
of particles, at least 
for the high wave numbers needed for the calculation of density 
fluctuation, appears quite probable. Even if the collision rates
among the particles are 
too small to produce such a state, there could be another mechanism at work.
As pointed out by Weinberg \cite{wein}, the strong gravitational interaction
at very early times would bring about thermal equilibrium, which, as we shall
show, could be maintained at least till the beginning of inflation.

The other point we discuss is the time at which quantum fluctuation must be
evaluated. It relates to the applicability of quantum field theory in curved 
space-time. As emphasised by DeWitt \cite{dewitt}, conventional quantum field 
theory requires the mode functions to be oscillatory in time, allowing 
positive and negative frequencies to be identified. While on flat 
space-time such modes naturally arise for field theories describing 
physical particles, their existence is not guaranteed on space-times 
with non-zero curvature. 
The reason is that the curvature gives rise to a damping-like term in the 
equation of motion for the mode functions. In the inflationary period this 
makes a  mode
oscillatory or damped, according as 
the associated physical wavelength is smaller or bigger  than  
the Hubble length. During this period physical wavelengths grow 
at a tremendous rate, while the Hubble length remains constant or 
approximately so. Thus even a wavelength lying initially deep inside the Hubble 
length would eventually go outside this length. So this time of exit  
marks the latest time at which we can evaluate the quantum fluctuation 
belonging 
to that particular wavelength.

In the literature, however, quantum fluctuations are actually evaluated at 
a time, when the modes have evolved well outside the the 
Hubble length, so as to be frozen \cite{copeland}. 
Of course, the complete problem of predicting the density fluctuation at 
the time of Hubble length reentry in a later radiation or matter 
dominated phase does involve the evolution of the fluctuation over a much 
longer period of time. But the question 
at hand is where the quantum fluctuation can be evaluated reliably.

In Sect. II we review the derivation of the thermal scalar propagator in the 
early universe. In particular, we 
show the behaviour of mode functions as the wavelengths grow and
discuss the validity of the assumed initial thermal equilibrium state.
In Sect. III we write the formula for density 
fluctuation in terms of the mode functions and discuss its dependence on the 
initial condition. In Sect. IV we consider the simple,
original model of extended inflation as an example \cite{mathia}.
Here we review the homogeneous classical solution for the inflaton field and
set up quantum theory in this background.
Finally, our concluding remarks are contained in 
Sect. V.

\section{Finite Temperature Scalar Propagator}
\setcounter{equation}{0}
\renewcommand{\theequation}{2.\arabic{equation}}

Consider a region of space in the early universe well within the
causal horizon. It can then be taken to be homogeneous and isotropic, 
admitting the (spatially flat) line element, 
\be ds^2=dt^2 - a^2(t)d{\vec x}^2~, \ee
where the scale factor $a(t)$ describes the expansion of the region of 
the universe. It constitutes the background space-time, which is perturbed by
quantum fluctuations. 
The action for the scalar field in this space-time may be generally
written as, \be S_{\phi}
={1\over 2}\int d^3xdt a^3(t)\left\{{\dot{\phi}}^2-{1\over a^2(t)}(\nabla 
\phi)^2-\mu^2(t)\phi^2-\lambda_1(t)\phi^3-\lambda_2(t)\phi^4+\cdots\right\}.\ee
Here we have already shifted the scalar field by the classical, homogeneous
field. 
The dots indicate interaction terms, if any, of the scalar field with other 
(gauge and matter) fields. 

We assume the different species of particles to be in thermal equilibrium
around some initial time $t_0$, which we conveniently take to be the time of 
transition of 
the radiation dominated phase to the inflationary phase. In
particular, the scalar particles belonging to $\phi(x)$ are also assumed to be
in thermal equilibrium. (This assumption will be examined at the end of this 
section.) 
The density matrix is then given by \be \rho = e^{-\beta_0 {\cal H}(t_0)}/Tr 
e^{-\beta_0 {\cal H}(t_0)},\ee where ${1/ \beta_0}=T(t_0)$ is the 
temperature at time $t_0$. 
The explicit time dependence of the Hamiltonian ${\cal H}(t)$ arises from that
of the scale factor and the homogeneous classical field. Note 
that the density matrix is constant in the Heisenberg representation.
Thus once the system is in the thermal equilibrium state,
the thermal propagator 
continues to hold even when the system deviates from this state.
 
To describe the time evolution of the system, it is most convenient to use
the real time formulation of thermal field theory \cite{Niemi}. In the
context of the early universe, the time path $C$ 
in the action integral consists of three segments
as shown in Fig.1 \cite{semenoff}. The points on it may be labelled by a
complex parameter $\tau$ such that
\be \tau = \left\{ \begin{array}{cc}t,\qquad & {\rm{on}}~ C_1~
{\rm{and}}~ C_2\\t_0-i
t,\qquad & {\rm{on}}~ C_3~. \end{array} \right.\ee
 

The action in the path integral corresponding to the segments $C_1$ and
$C_2$ is in Minkowski space, while it is Euclidean on $C_3$. It should be
noted, however, that the scale factor is not continued to Euclidean space on
$C_3$ : since the Hamiltonian for the segment is ${\cal H}(t_0)$, 
the scale factor
remains fixed at $t_0$. 

We now review the derivation of the thermal propagator \cite{semenoff},
\cite{leutwyler}. After a partial integration, the quadratic terms in
$S_{\phi}$ becomes \cite{comment1},
\be S_0= -{1\over 2}\int_C
d\tau \int d^3x
\phi D\phi ~+ ~\rm{boundary ~terms}, \ee where
\be D=\left\{\begin{array}{ccc}a^3\left({d^2\over d\tau^2}+3
{{\dot a\over a}}{d\over d\tau}+\omega^2\right),\qquad &\omega^2(\tau)=
 -{1\over a^2(\tau)}\nabla^2+\mu^2(\tau)~,~ &\tau\in C_1,C_2\\a_0^3
\left({d^2\over d\tau^2}+\omega_0^2\right),\qquad &\omega_0^2=
-{1\over a_0^2}\nabla^2 +\mu^2(t_0)~,&~\tau\in~C_3~. \end{array} \right. \ee
We use the abbreviations, $a_0=a(t_0),~ \omega_0 =\omega (t_0),~ T_0 =T(t_0).$
For the boundary terms to vanish, it is necessary that both $\phi$ and 
${d\phi\over d\tau}$ match at the joining of the segments $C_1$ and $C_2$,
of $C_2$ and $C_3$ as well as at the free ends of $C_3$ and $C_1$.

The (time ordered) thermal propagator $Tr \rho T\phi(x) \phi(x^{'})$ will
 be denoted by
$G_{\beta}(x,\tau ;x^{'},\tau ^{'})$ or $ <T \phi(x)\phi(x^{'})>$. 
It satisfies \be D G_{\beta}({\vec x},\tau ;{\vec x}^{'},
\tau^{'})=-i\delta^3({\vec x}-
{\vec x}^{'})
\delta(\tau -\tau^{'})~, \ee with boundary conditions 
following from the matching 
of $\phi$ and ${d\phi\over d\tau}$ mentioned above. For the spatial 
Fourier transform of the propagator, defined by 
\be G_{\beta}({\vec x},\tau, \tau^{'})~= \int {d^3k\over (2\pi)^3} e^{i{\vec k}
\cdot{\vec x}} G_{\beta}({\vec k}, \tau, \tau^{'}), \ee 
it reduces to
\be D G_{\beta}({\vec k};\tau , \tau^{'})=
-i\delta(\tau -\tau^{'}), \ee 
where $-\nabla^2$ appearing in the expressions for $\omega^2(\tau)$ and 
$\omega_0^2$ is to be replaced now by $k^2$.

To construct the propagator we first find the mode functions.  
On the contour $C_3$ they satisfy \be \left({d^2\over
 d\tau^2}+\omega_0^2\right)h^{\pm}(\tau)=0, \ee 
giving
\be h^{\pm}(\tau)={1\over \sqrt{2\omega_0 a_0^3}}
e^{\mp i\omega_0\tau}, \qquad \tau=t_0-it
~ \in C_3. \ee
The normalization satisfies the Wronskian condition, ${\dot h}^+(\tau)h^-(\tau)
-{\dot h}^-(\tau)h^+(\tau)=-i/ a_0^3.$
The mode functions on the real segments $C_1$ and $C_2$ are the solutions of 
\be \left(
{d^2\over d\tau^2}+3{{\dot a}\over a}{d\over d\tau}+\omega^2(\tau)
\right) g^{\pm}(\tau)
=0, \qquad \tau=t ~\in C_{1,2}~,\ee
with normalisation fixed again by the Wronskian condition, 
${\dot g}^+(\tau)g^-(\tau)-{\dot g}^-(\tau)g^+(\tau)=-i/ a^3(\tau).$
To see the nature of these solutions, we put \be g^{\pm}(\tau)=
 a^{-3/2}
{\bar g}^{\pm}(\tau),\ee where ${\bar g}$ satisfies \be \left({d^2\over 
d\tau^2}+\bar{\omega}^2(\tau)\right){\bar g}^{\pm}(\tau)=0~, \ee with 
\be \bar{\omega}^2(t)={k^2\over a^2}+\mu^2-{9\over 4}\left( H^2+{2\over 3}
{\dot H}\right),\qquad H(t)={{\dot a}(t)\over a(t)}~.
\ee
For a power law behaviour of the scale factor, $a(t)~\sim t^p$, it becomes
\be {\bar\omega}^2(t)={k^2\over a^2}+\mu^2-{9\over 4}\left(1-{2\over 3p}
\right)H^2~.
\ee

It is now simple to identify the modes, which belong to conventional quantum
field theory. The magnitude of $ \mu (t) $ is usually small 
compared to $H(t)$. Thus in the radiation dominated phase $(p={1\over 2})$, 
${\bar\omega}^2$ is positive for all values of $k$ and it may be possible 
to define oscillatory modes belonging to positive and negative frequencies, 
at least in a quasi-static way. We show below that this is indeed the case
around the time $t_0$, when we can solve (2.14) in the JWKB approximation
to get \cite{comment2},
\be {\bar g}^{\pm}(\tau)={1\over \sqrt{ 2{\bar\omega}(\tau)}} e^{\mp
i\int_{t_0}^{\tau}d\tau^{'}{\bar\omega}(\tau^{'})},\qquad \tau\simeq t_0.
\ee
We thus have a valid quantum field theory around the time $t_0$.

But in the inflationary phase $(p~\gg~1)$, a mode is oscillatory only if its
physical wavelength $2\pi a(t)/k$ is small compared to the Hubble length
$H^{-1} (t)$. As
inflation progresses, the scale factor increases enormously, while $H(t)$ is
approximately constant. Thus more and more modes go out of Hubble length and
behave as damped waves, having no interpretation in quantum field theory.

The solutions $g^{\pm}(\tau)$ and $h^{\pm}(\tau)$ may now be joined to
form the functions $f^{\pm}(\tau)$ on the entire contour $C$,
\be f^{\pm}(\tau)= \left\{ \begin{array}{c} g^{\pm}(\tau),~~\tau~\in 
~C_{1,2}
\\  h^{\pm}(\tau), ~~\tau~\in~ C_3~. \end{array} \right .\ee
By definition, $f^{\pm}(\tau)$ obey the continuity conditions relating the
segments $C_1$ and $C_2$. Using Eqs (2.11) and (2.17), we see that
the conditions connecting $C_2$ and $C_3$ are 
also well satisfied if $H(t_0)$ is small 
compared to $k/a(t_0)$ \cite{comment3}.
A particular solution to (2.9) may now be written as \be G_0(k:\tau,
\tau^{'}) = f^+(\tau)f^-(\tau^{'})\theta_c(\tau-\tau^{'})+f^+(\tau^{'})
f^-(\tau)\theta_c(\tau^{'}-\tau),\ee where $\theta_c$ is a step function on 
the contour. It satisfies the continuity conditions at the junctions of 
segments $C_1$ and $C_2$ as well as of $C_2$ and $C_3$, because $f^{\pm}
(\tau)$ does it. To satisfy the remaining (thermal) 
continuity condition at the ends of $C_1$ and $C_3$, we add to it the most 
general solution of the homogeneous equation, \be G_{\beta}(k;\tau,\tau^{'})=
G_0(k,\tau,\tau^{'})+\sum_{i,j=1}^2 f^i(\tau)\Lambda^{ij}f^j(\tau^{'})~.\ee
The superscript $(\pm)$ on the mode functions are replaced 
temporarily by 1 and 2 to 
use matrix notation. The $2\times 2$ constant coefficient matrix $\Lambda$
is uniquely determined by the thermal conditions \cite{comment4}. We get 
\be G_{\beta}(k,
\tau,\tau^{'})=f^+(\tau)f^-(\tau^{'})\{\theta_c (\tau -\tau^{'})+
n(\omega_0)\}+f^-(\tau)
f^+(\tau^{'})\{\theta_c(\tau^{'}-\tau )+n(\omega_0) \}~, \ee 
where $n(\omega_0)$ is the bosonic distribution function 
\be
n(\omega_0)=(e^{\beta_0\omega_0}-1)^{-1}.
\ee

For tree level calculations we need the Green's function only
on the real axis $C_1$. 
Writing henceforth $\tau=t$, this is given by \be \langle\phi({\vec x},t)
\phi({\vec {x^{'}}},t^{'})\rangle =\int {d^3k\over (2\pi)^3}
e^{i{\vec k}.({\vec x}-
{\vec x}^{'})}
(1+n(\omega_0)) g^+(t)g^-(t^{'})~, ~~ t > t^{'}~, \ee 
where the mode functions $g^{\pm}(t)$ are solutions of (2.12).

We now come back to the assumption of the initial thermal equilibrium state. 
Such an initial state can be ensured in an expanding universe if collisions 
among particles 
occur at a rate faster than the expansion rate of the universe. While 
this condition holds for species interacting through (relatively large) gauge 
coupling, it may not hold for particles of the inflaton field, which is a gauge
singlet and has weak self-interaction. We discuss below the other mechanism,
mentioned in the introduction, which could give rise to 
thermal equilibrium around the time $t_0$.

At the Planck time $t_P$, the strong gravitational interaction brings all
species into thermal equilibrium \cite{wein}. 
If the system is quantised at this time in a cubic volume with sides small 
compared to the Hubble
length, the longest wavelength will be well inside this length, ${\it i.e.}$
\be {k\over a(t_P)}~>~ \pi H(t_P),\ee even for the smallest wavenumber.
Then eq (2.16) simplifies to 
\be {\bar \omega}(t_P)={k\over a(t_P)},
\ee
to a good approximation and the density distribution becomes 
\be n(\omega(t_P))={1\over e^{k/a(t_P)
T(t_P)} -1}~.
\ee
We now bring the inequality (2.24) to the time $t_0$ ,
\be
{k\over a(t_0)}~>~ {a(t_P)\over {a(t_0)}}{H(t_P)\over {H(t_0)}}\pi H(t_0)~
={m_P\over {T_0}} \pi H(t_0),
\ee
where $m_P$ is the Planck mass. In the last step we have used the radiation
dominated solution for $a(t)$. (See eqns. (4.6-7) below.) 
The temperature $T_0$ is 
given by the grand unification scale, $T_0 \sim 10^{15}~ GeV$, so 
that $m_P/T_0~ \sim 10^5$. Thus the relation (2.25) at
time $t_P$ continues to hold throughout the radiation dominated phase; in fact,
it becomes more and more accurate as $t$ increases from $t_P$. Clearly the
equilibrium distribution (2.26) established at time $t_P$ is well maintained 
at least till the time $t_0$.

\section{Density fluctuation formula}
\setcounter{equation}{0}
\renewcommand{\theequation}{3.\arabic{equation}}

The density inhomogeneity (at time $t$) is measured by the mean square 
fluctuation in the density function $\rho({\vec x},t)$ \cite{kolb}
\cite{padma}, \be \left({\delta
\rho\over \rho}\right)^2 =~\left\langle \left({\rho({\vec x},t)-{\bar\rho}
(t)\over {\bar
\rho}(t)}\right)^2\right\rangle_x ,\ee where $\langle\cdots\rangle_x$ 
denotes averaging over space and 
${\bar\rho}$ is the homogeneous background density, ${\bar\rho} = \langle
 \rho
({\vec x},t)\rangle_x$. In the inflationary scenario, 
this fluctuation in the early 
universe is supposed to arise from quantum fluctuation. 
We may calculate the latter by evaluating an expression 
similar to (3.1), with $\rho({\vec x},t)$ replaced by 
the corresponding operator $\hat{\rho}({\vec x},t)$ and the averaging process
by taking the expectation value in an appropriate state. 
There is, however, a  technical problem with this quantum 
version, as it involves the product of $\hat{\rho}({\vec x},t)$ 
with itself at the same 
space-time point, which is not defined in quantum field theory. The problem may be avoided by taking the smeared density function \cite{gupi}, 
\be \rho_l({\vec x},t)=N
\int d^3ye^{-y^2/2l^2}\rho({\vec x}+{\vec y},t), \ee where $l$ is an arbitrary
smearing length and $N$ an irrelevant normalization factor. The fluctuation
in $\rho_l$ is given by 
\be \left({\delta\rho_l\over \rho_l}\right)^2_c=~
\left\langle\left({ \rho_l({\vec x},t)-{\bar\rho}_l(t)\over {\bar\rho}_l(t)}
\right)^2\right\rangle_x~, \ee where the subscript $c$ stands for classical.
The corresponding quantum fluctuation, denoted by the subscript $q$, 
is now well 
defined, \be \left({\delta\rho_l\over \rho_l}\right)^2_q = ~ 
\left\langle\left({{\hat{\rho_l}({\vec x}
,t)}-{\bar\rho (t)}\over {\bar\rho}(t)}\right)^2\right\rangle~,\ee
where $\langle\cdots\rangle$ stands for the expectation 
value in the initial thermal state
defined by eqn (2.3).
 
To treat perturbation on different length scales, one writes \be \rho({\vec x}
,t)= {\bar\rho}(t)(1+\delta({\vec x},t)),\ee and Fourier analyzes the so-called 
density contrast, $\delta({\vec x},t),$ \be \delta({\vec x},t)={1\over \sqrt V}
\sum_k\delta_k(t)e^{i{\vec k}.{\vec x}}, \ee where $V$ is a volume within 
the Hubble length to begin with. In the limit of large volume, (3.3) 
becomes \be \left({\delta\rho_l\over \rho_l}\right)^2_c=
\int{d^3k\over (2\pi)^3}|\delta_k(t)|^2e^{-k^2l^2}. \ee 

The energy density operator ${\hat\rho}({\vec x},t)$ is the 
time-time component of energy 
momentum tensor, \be T_{\mu\nu}=\partial_\mu\Psi\partial_\nu\Psi-g_{\mu\nu}
\left({1\over 2}g^{\mu\nu}\partial_\alpha\Psi\partial_\beta\Psi -V(\Psi)
\right),\ee for the full scalar field $\Psi(x)$. Here the potential function 
$V(\Psi)$ depends on the model considered. We  shift $\Psi(x)$ by the 
homogeneous classical field $\psi(x),$ \be \Psi({\vec x},t)=\psi(t)+\phi(
{\vec x},t), \ee such that for the quantum field $<\phi(x)>~=0$. The ${\hat\rho}
(x)$ is given by \be \hat\rho(x)=\bar\rho(t)+{\hat U}(x), \ee 
where \be \bar\rho(t)={1\over 2}{\dot\psi}^2+V(\psi),\ee 
and \be {\hat U}(x)=r(t)\phi(x)+
s(t) {\dot\phi}(x), \ee to first order in $\phi(x)$. The coefficients $r(t), s(t)$ 
in (3.12) depend on the classical field and other parameters in the 
potential $V(\Psi).$ Terms in ${\hat U}$, 
which are of higher order in $\phi$ would give loop contribution to the density
fluctuation and are neglected. Using (3.10), (3.12) and (2.23), the
expectation value in (3.4) may be evaluated to give, 
\br \left({\delta\rho_l
\over \rho_l}\right)^2_q &=& {1 \over {\bar\rho}^2(t)}\int d^3x d^3y 
e^{-(x^2+y^2)/
l^2}<{\hat U}(x,t){\hat U}(y,t)> \nonumber\\&=& {1 \over {\bar\rho}^2(t)}
\int {d^3k\over (2\pi)^3}
(1+n(\omega_0))\mid r(t)g_k(t)+s(t){\dot g}_k(t)\mid ^2 e^{-k^2l^2}~. \er
Comparing eqs (3.7) and (3.13) we get the desired result, \be \mid\delta_k(t)\mid^2={1 \over 
{\bar\rho}^2}(1+n(\omega_0))\mid r(t)g_k(t)+s(t){\dot g}_k(t)\mid^2 .\ee
Following our discussion in Sect. II, we evaluate it at time $t_h$, when the
wavelength crosses the Hubble length, 
\[ {k\over a(t_h)} = 2\pi H(t_h) .\] 
The treatment of the 
evolution of density perturbation outside the Hubble length 
is standard \cite{kolb}, \cite{padma}. At the time of its reentry within the
Hubble radius of the post-inflationary radiation dominated epoch,
it is given by \cite{mallik}, 
\be \left({\delta\rho \over \rho}
\right)_H
={\sqrt{k^3(1+n(\omega_0))}\mid rg_k(t_h)+s{\dot g}_k(t_h)\mid 
\over 3{\sqrt 2} \pi ({\bar\rho} +
{\bar p})_{t_h}}~,\ee where ${\bar p}$ is the homogeneous pressure, 
${\bar p}={1\over 2}{\dot \psi}^2-V(\psi).$ 

It is simple to estimate $n(\omega_0)$ in the range of
$k/a_0$, which is of interest. We write \be {k\over a_0}={k\over a(t_p)}\cdot
{a(t_p)\over a(t_e)}\cdot{a(t_e)\over a(t_0)}~. \ee 
From the time $t_e$ when the 
inflation ends till the present time $t_p$, the universe expands adiabatically,
so that $a(t_p)/a(t_e)\simeq T_0/T_p$. The other ratio $a(t_e)/a(t_0)
\equiv Z$ gives the magnitude of inflation. We thus get 
\be {k\over a_0T_0}= {2\pi\over 
\lambda(t_p)}{Z\over T_p}~\sim {1\over \lambda_{Mpc}}\cdot {Z\over 10^{25}}~, 
\ee where $T_p=2.7 K$ and $\lambda_{Mpc}$ is $\lambda(t_p)$ expressed in $Mpc$.
The wavelengths of interest stretch over the range $1<\lambda_{Mpc}<10^4$. 
To solve the problems of the 
standard cosmology we need $Z>10^{25}$. But actually $Z$ exceeds this 
limit by many orders of magnitude in most of the models of inflation. 
Thus $k/a_0T_0$ is large in these models and we may set $n(\omega_0)=0$ 
in the expression (3.14).

We thus see that although the initial thermal equilibrium state does
produce a factor in 
the expression for the density inhomogeneity, its magnitude turns out to be 
unity, justifying the use of zero temperature propagator for 
its evaluation. Nevertheless it is important to know the initial state, as
there are other quantities, such as the duration of inflation, which may
depend sensitively on it.

\section{Example of extended inflation}
\setcounter{equation}{0}
\renewcommand{\theequation}{4.\arabic{equation}}

In this section we consider the well-studied, original model of extended 
inflation \cite{mathia}. We first summarise, for the sake of completeness, 
the classical inflationary solution for the scalar field, which joins 
smoothly to the radiation dominated solutions before and after the inflation. 
We then write the Lagrangian for the quantum field in the background 
of this homogeneous field. The mode functions and the (vacuum) propagator  
has already been evaluated in a recent work \cite{mallik}. Here we only
discuss the possibility of establishing thermal equilibrium through collisions.

The model is based on the Brans-
Dicke (BD) theory of gravity given by the action \cite{brans},
\be S=\int d^4x {\sqrt g}\left({R\over 16\pi }\Phi+{\omega\over 16\pi}g^{\mu 
\nu}{ \partial_\mu\Phi\partial_\nu\Phi\over \Phi} +
{\cal L}_{matter}\right)~ .\ee
The time dependence of the BD field $\Phi$ makes the effective gravitational
`constant' vary with time. $\omega$ is a dimensionless parameter. 
${\cal L}_{matter}$ represents the contribution of
all other fields including the inflaton field $\sigma$, \be {\cal L}_{matter}
={1\over 2}g^{\mu \nu}\partial_{\mu}\sigma\partial_{\nu}\sigma- V(\sigma)+
\cdots ~ .\ee 
This simple model is not realistic, however. The bubble nucleation 
problem can be solved for $\omega \leq 25$, while astrophysical observation
constrains it to $\omega > 500$. Our purpose here is to  examine in this toy 
model  the possibility of realising thermal equilibrium as the initial state
for the inflationary epoch. In the following we take $\omega > 25$, say.

The equations satisfied by the classical fields, ${\it viz.}$ the scale 
factor $a(t)$ and the homogeneous part $\varphi(t)$ of the BD field 
$\Phi(x)$ are \cite{mathia},
\be \ddot{\varphi}+
3H\dot{\varphi}={8\pi\over 2\omega+3}(\rho -3p)~~, \qquad H={{\dot a}\over 
a},\ee \be H^2={8\pi\rho\over 3{\varphi}}+{\omega\over 6}\left({\dot{\varphi}
\over {\varphi}}\right)^2- H\left({\dot{\varphi}\over {\varphi}}\right)~,\ee
where the energy density $\rho$ and the pressure $p$ are generated by 
${\cal L}_{matter}$. If the masses are small compared to the temperature, 
they are \be \rho={\pi^2\over 30}NT^4+M^4, \qquad p={\pi^2\over 90}NT^4
-M^4, \ee where $M^4 = V(0)$ is the false vacuum energy density and $N$ counts 
the total number of effective degrees of freedom.

In the pre-inflationary radiation dominated epoch, $\varphi(t)$ is 
constant \cite{wein2},
\be \varphi (t)=\varphi(t_0)~, \qquad a(t)\propto {\sqrt t}, \ee 
giving \be T^2t=\sqrt{{
45\varphi(t_0)\over 16\pi^3N}}, \ee while in the inflationary epoch the 
solutions may be written as \cite{mathia} 
\be \varphi(t)= \varphi(t_0)\{1+B(t-t_0)\}^2,\ee
with $$ B{\sqrt \varphi(t_0)}={M^2\over \omega q},\qquad 
q={\sqrt{(6\omega+5)(2\omega+3)\over 32\pi \omega^2}}~,$$ and
\be a(t)=a(t_0)\{1+B(t-t_0)\}^{\omega +1/2}~. \ee 
The two 
sets of solutions join at $t_0$, when we have approximately 
\be {\pi^2\over 30}NT_0^4\simeq M^4, \ee giving $T_0\simeq M$ and \be T_0^2
t_0=\sqrt{{45\varphi(t_0)\over 16\pi^3 N}}~.
\ee Equating the values of $H(t_0)$
for the two solutions, we get another relation \be {1\over 2 t_0}=(\omega +
{1\over 2}) B~. \ee It is no new constraint, coinciding with (4.11) to 
leading order in $\omega$.

We now assume as usual that the inflationary period ends 
instantly at time $t_e$ in
radiation domination again at a temperature $T\sim T_0$. 
Then $\varphi(t)$ ceases 
to vary appreciably, so that we have \be \varphi(t)=\varphi(t_e)=m_P^2,
\qquad t\geq t_e~, \ee where $m_P\equiv G^{-1/2}$ is the present value of the 
Planck mass.  Using (4.8) and (4.9) we can relate $\varphi (t_o)$ 
to the amount of inflation $Z$,\be \varphi(t_0)=m_P^2Z^{-{4/(2\omega+1)}}
,\qquad
Z={a(t_e)\over a(t_0)}~.\ee 

To set up a semi-classical quantum theory in the background of the above
solutions, one faces the problem of dealing with the non-standard form of
the action (4.1). This problem may be avoided by going over to a new 
( Einstein ) frame from the present ( Jordan ) frame by a conformal transformation of the metric. Denoting quantities in the new frame by a bar, the required 
transformation is \cite{holman}, 
\be {\bar g}_{\mu \nu}({\vec x},t)=\Omega^{-2}(t)
g_{\mu \nu}({\vec x},t), \ee where \be \Omega^2(t)={m_P^2\over \Phi(t)}~.\ee
One also introduces a field $\Psi$ to replace $\Phi$, \be \Psi=\chi~ {\rm{ln}}
\left({\Phi\over m_P^2}\right), \qquad \chi=\sqrt{{2\omega+3\over 16\pi}}
m_P, \ee which brings the kinetic term for $\Phi$ in the canonical form. 
Assuming $\sigma =$ constant in the inflationary phase, the new action
becomes, \be {\bar S}=\int d^4x\sqrt{\bar g}\left\{{{\bar R}\over 16\pi m_P^2}
+{1\over 2}{\bar g}^{\mu \nu}\partial_{\mu}\Psi\partial_{\nu}
\Psi+V(\Psi)\right\},\ee
where $V(\Psi)=M^4e^{-2\Psi /\chi}$. In the Einstein frame $\Psi$ plays 
the role of the inflaton field.

The new metric may be brought back to the Robertson-Walker form by a 
redefinition of the time coordinate \cite{holman},
\be d{\bar t}^2-{\bar a}^2({\bar t})
d{\vec x}^2= \Omega ^{-2}(t)\{dt^2-a^2(t)d{\vec x}^2\}~, \ee giving \be 
d{\bar t} = \Omega^{-1}(t)dt, \qquad {\bar a}({\bar t})= \Omega^{-1}(t)
a(t).\ee  Using (4.16) and (4.20), the inflationary solutions 
(4.8-9) in the Jordan frame may be recast in the Einstein frame. Integrating
the first equation in (4.20) and choosing the constant of integration properly, we get
\cite{guja},
\be 1+C({\bar t}-{\bar t}_0) =\{1+B(t-t_0)\}^2, 
\qquad C={2m_PB\over \sqrt{ \varphi(t_0)}}~
.\ee Then the new scale factor is obtained from the second equation in
(4.20), \be {\bar a}
({\bar t})={\bar a}({\bar t}_0)\{1+C({\bar t}-{\bar t}_0)\}^{(2\omega+3)/ 4},
\qquad {\bar a}({\bar t}_0)=a(t_0){\sqrt{\varphi(t_0)}\over m_P}~ .\ee
The classical homogeneous part of $\Psi$ is obtained from (4.8) and (4.17),
 \be\psi({\bar t})= \psi ({\bar t}_0)+\chi 
{\rm{ln}} \{1+C({\bar t}
-{\bar t}_0)\}, \qquad 
\psi({\bar t}_0)=\chi {\rm{ln}}\left({\varphi(t_0)\over m_P^2}\right). \ee
These solutions could, of course, have been found by solving the equations 
obtained by varying the action (4.18) in the Einstein frame.

The free part of the action (4.18) is now in the standard form for 
developing quantum field theory. We introduce the quantum field $\phi
(x)$ by decomposing $\Psi$ as, \be \Psi({\vec x},t)=\psi(t)+\phi({\vec x},t).\ee
The exponential potential $V(\Psi)$ as such has no place among 
renormalizable quantum field theories. But we can treat it as an 
effective potential and retain 
terms up to fourth power in $\phi$. The action for $\phi$ then becomes 
\be {\bar S}=\int d^3x d{\bar t} {\bar a}^3\left\{{1\over 2}\left({d\phi\over 
d{\bar t}}\right)^2-
{1\over 2{\bar a}^2}(\nabla\phi)^2-{1\over 2}\mu ^2\phi^2-\lambda_1\phi^3-
\lambda_2\phi^4\right\}~,\ee where 
\br \mu^2({\bar t})&\simeq&{12\over \omega}{\bar H}^2({\bar t})~,\\\nonumber
\lambda_1({\bar t})&\simeq& -{8{\sqrt{2\pi}}\over \omega^{3/2}}
{{\bar H}^2({\bar t})\over m_P}~,\\
\nonumber \lambda_2({\bar t})&\simeq&{16\pi\over \omega^2}{{\bar H}^2({\bar t})
\over m_P^2}~.\er 

The magnitudes of $\mu^2,~\lambda_1$ and $\lambda_2$ depend on
$1/ \varphi(t_0)$, the effective gravitational constant in the
pre-inflationary, radiation dominated phase. The relation (4.14) is not
useful in determining $\varphi(t_0)$ as $Z$ itself depends on the 
details of the models of inflation. However a lower bound on $\varphi(t_0)$
may be obtained from an earlier requirement that $\mu(t_0)~<~ T_0$, giving
$$\varphi(t_0)~>~{\sqrt {32\pi\over \omega}}\cdot Mm_P. $$ 
From the corresponding bounds on $\lambda_1$ and $\lambda_2$, it is simple to
find that the collision rate, $\Gamma_{coll}~< ~ \left({4\pi\over 3\omega}
\right)^2 \left({M\over m_P}\right)^4 M~,$ which is hopelessly small compared
to the expansion rate, $\Gamma_{exp}={\bar H}({\bar t}_0)$. Even though 
collisions are totally ineffective to maintain thermal equilibrium, we have
shown in Sect. II that
gravitational interaction at very early times would ensure equilibrium 
distribution of the $\varphi$ particles around the time $t_0$.

There are several earlier calculations of density 
fluctuation in this model \cite{guja}, \cite{kst}. All are 
based on the result for the vacuum propagator for the scalar field, which we 
justify in Sect. III. However earlier works generally evaluate the quantum 
fluctuation outside the Hubble length, which following our discussion in 
Sect. II, cannot be interpreted properly in quantum field theory. In a 
recent work \cite{mallik}, we evaluate this density fluctuation at the time
of Hubble length exit using the formula (3.15). We find a result, which is an 
order of magnitude bigger than the earlier ones.

\section{Conclusion}
\setcounter{equation}{0}
\renewcommand{\theequation}{5.\arabic{equation}}

In the present work we assume the inflationary epoch to begin in a 
state of thermal equilibrium and study its effect on the quantum fluctuation 
in the energy density calculated during this epoch. This
initial state including the scalar particles appears 
quite likely even if their self -interaction is too feeble to ensure it.
We show that the thermal equilibrium established at very early times through
the-then strong gravitational interaction would be maintained till 
the beginning of inflation. 
By evaluating the scalar field propagator with thermal boundary conditions,
we find a result for the
density fluctuation, which 
differs from the one calculated with the vacuum 
propagator by the factor $\{1+(e^{k_0/a_0T_0}-1)^{-1}\}$.
Clearly the factor does not go to unity as time passes but is a constant
depending on the physical wave number and the temperature referred to 
the initial time $t_0$.  

It turns out, however, that for wave numbers of interest in the present 
universe, this factor is unity in models where the amount of inflation 
exceeds by many orders of magnitude the minimal amount required to solve 
the problems of standard cosmology. Thus numerically
the calculation of fluctuation in the vacuum state is justified.

We also point out that the conventional quantum field theory applies on
curved space-time as long as the modes oscillate. This requires that we
evaluate the quantum fluctuation, at the latest, when the corresponding 
wavelength crosses the
Hubble length. Previous works \cite{copeland}, 
however, evaluate it as a rule for wavelengths well
outside this length, where the modes freeze. As we have shown in a recent
work \cite{mallik}, this difference in the calculation leads to an 
increase in the result
by a factor of five for the model of extended inflation. 

Finally we note that as long as the modes are within the Hubble length,
they retain a thermal equilibrium distribution. Thus although the initial 
state of thermal equilibrium comprising all modes is not maintained during
inflation, the modes
relevant for the calculation of density fluctuation are those still
in an equilibrium distribution.

\bigskip

{\bf {Acknowledgement}}\\
One of us ( S. M.) would like to thank Dr. S. Sarkar for his encouragement.


\begin{center}
\begin{picture}(300,300)(0,0)
\ArrowLine(0,0)(250,-7) \Text(100,10)[]{$C_1$} \Text(-10,10)[]{$t_0$}
\ArrowLine(250,-7)(0,-14) \Text(100,-24)[]{$C_2$} 
\Text(-30,-14)[]{$t_0-i\epsilon$}
\ArrowLine(0,-14)(0,-100) \Text(-12,-60)[]{$C_3$} 
\Text(0,-110)[]{$t_0-i\beta$}
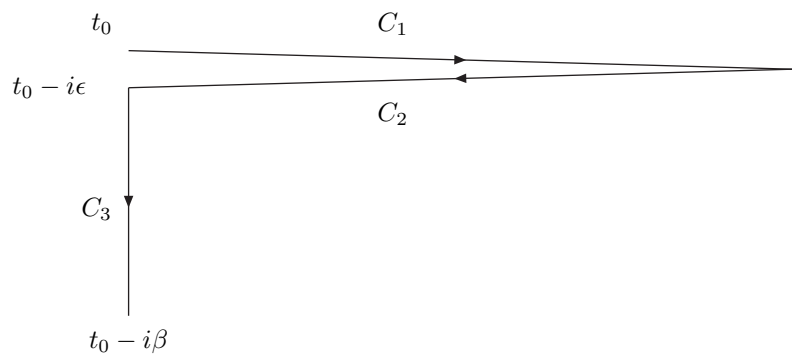
\Text(30,-150)[lb]{Fig. 1. Time path of real time thermal field theory}
\end{picture}
\end{center}


\begin{thebibliography}{99}
\bibitem{comment1} The quadratic action is generally improved by replacing 
$\mu(t)$ with the effective mass the scalar 
particle acquires at finite temperature 
through the self- and gauge-interactions. But for the 
inflaton field, which has no gauge interaction and is only weakly self-coupled, 
this modification is not significant.
\bibitem{comment3} Let us mention here the problem associated with the
density matrix (2.3) defined sharply at time $t_0$.
This procedure of instant thermalisation
leads to additional short distance singularities in the propagator, 
not present at zero temperature, making the theory non-renormalisable.  
Here we thermalize the system in a static background with constant scale
factor prior to $t_0$ and then connect it smoothly to the actual scale factor.
The resulting thermal propagator does not depend on the details of the 
interpolating scale factor, if the condition $k/a(t_0)~\gg~  H(t_0)$ 
is satisfied. See ref. \cite{leutwyler} for more details.
\bibitem{comment4}  It is interesting to note 
that although we insisted on oscillatory modes with positive and negative
frequencies to define quantum field theory, the propagator (2.20) 
is invariant
under a reparametrization mixing the mode functions, 
$ f^+(\tau)\rightarrow \alpha
f^+(\tau)+\beta f^-(\tau)~,~ f^-(\tau)\rightarrow \gamma f^+
(\tau)+\delta 
f^-(\tau),$  the constant coefficients $\alpha,\beta,\gamma$ and $\delta$
satisfying $\alpha\delta - \beta\gamma =1$. It is a property of the thermal
propagator also on flat space-time.
\end{thebibliography}
\end{document}